\documentclass{aa}  

\usepackage[switch, pagewise, mathlines, displaymath]{lineno}

\usepackage{graphicx}
\usepackage{txfonts}
\usepackage{hyperref}

\begin{document} 


\title{A spatial likelihood analysis for MAGIC telescope data}

\subtitle{From instrument response modelling to spectral extraction}

\author{Ie. Vovk\inst{1}        
        \and
        M. Strzys\inst{1}
        \and
        C. Fruck\inst{2,1}
        }

\institute{Max-Planck-Institut f\"ur Physik,
           F\"ohringer Ring 6, D-80805 Munich, Germany \and Technische Universit\"at M\"unchen, Physik-Department, James-Franck-Str. 1, D-85748 Garching\\
           \email{ie.vovk@mpp.mpg.de}
           \email{strzys@mpp.mpg.de}
           \email{fruck@mpp.mpg.de}
           }

\date{Received --; accepted --}

\abstract
 {
The increase in sensitivity of Imaging Atmospheric Cherenkov Telescopes (IACTs) has lead to numerous detections of extended $\gamma$-ray sources at TeV energies, sometimes of sizes comparable to the instrument's field of view (FoV). This creates a demand for advanced and flexible data analysis methods, able to extract source information by utilising the photon counts in the entire FoV.
 }
 {
 We present a new software package, ``SkyPrism'', aimed at performing 2D (3D if energy is considered) fits of IACT data, possibly containing multiple and extended sources, based on sky images binned in energy. Though the development of this package was focused on the analysis of data collected with the MAGIC telescopes, it can further be adapted to other instruments, such as the future Cherenkov Telescope Array (CTA).
 }
 {
 We have developed a set of tools that, apart from sky images (count maps), compute the instrument response functions (IRFs) of MAGIC (effective exposure throughout the FoV, point spread function (PSF), energy resolution and background shape), based on the input data, Monte-Carlo simulations and the pointing track of the telescopes. With this information, the presented package can perform a simultaneous maximum likelihood fit of source models of arbitrary morphology to the sky images providing energy spectra, detection significances, and upper limits.
 }
 {
 We demonstrate that the SkyPrism tool accurately reconstructs the MAGIC PSF, on and off-axis performance as well as the underlying background. We further show that for a point source analysis with MAGIC's default observational settings, SkyPrism gives results compatible with those of the standard tools while being more flexible and widely applicable.
 }
 {}

\keywords{      Gamma rays: general --
                Methods: data analysis --
                Physical data and processes --
                Techniques: image processing
               }

\maketitle

\section{Introduction}
\label{sect:introduction}

Imaging Atmospheric Cherenkov Telescopes (IACTs) observe astrophysical sources through the detection of optical light from Extended Atmospheric Showers (EASs), induced by $\gamma$-rays of energies $\gtrsim 10$~GeV. These observations are usually performed in the presence of a relatively high background from cosmic ray (CR) initiated EASs, which significantly outnumber the signal counts. Despite advances in background suppression techniques~\citep[e.g.][]{Magic_RF}, a non-negligible part of CR events survive the event selection criteria forming an irreducible, gamma-like background throughout the entire field of view of an IACT.

Traditionally, this remaining background is managed by defining a source region and (possibly several) background control regions \citep{berge_background_2007}. The source signal is extracted as the difference in the counts from the source and background area. Given the simplicity of this approach, it is also commonly used to extract the spectral information of astrophysical sources.

Such an ``aperture photometry'' method is also adopted in the standard data analysis routines of the MAGIC collaboration, operating one of the state-of-the-art IACTs, located at the Canary island of La Palma, Spain~\citep{Magic03, aleksic_major_2016-1, aleksic_major_2016}. Despite being successful at analysing point sources, this method has several drawbacks when analysing extended astrophysical objects:

\begin{itemize}
 \item the shape of the source/background subtraction regions must be adapted to the real source shape, otherwise it results in excessive background counts in the source region and consequently a loss of sensitivity.
 \item the signal of overlapping sources or source components -- whether nearby point-like or extended -- can not be easily distinguished.
\end{itemize}

The steady increase in sensitivity, due to improved technology or new instruments such as the upcoming Cherenkov Telescope Array ~\citep[CTA,][]{CTA_concept}, and thereby the detection of even fainter sources and source components, further intensifies the need for alternative analytical approaches.

A way to overcome these issues is using a 2D spacial modelling of the measured sky signal: by applying the known instrument response to an assumed source model, one can obtain an image of the model as it would be seen by the telescope. This model, together with the estimated background map, is fitted to the measured sky image to estimate the most likely flux of the model sources in the observed sky region. When combined with the fitting of the spectral energy distribution (SED) of the sources, this procedure becomes a 3D fit in the ``x-y-Energy'' phase space.

This approach has successfully been applied for the space-based EGRET ~\citep{mattox_likelihood_1996} and {\it Fermi}-LAT~\footnote{https://fermi.gsfc.nasa.gov/ssc/data/analysis/scitools/overview.html} missions. Furthermore it is considered for the analysis of CTA data in software packages such as CTools ~\citep{CTools} or Gammapy\footnote{https://github.com/gammapy/gammapy}~\citep{gammapy}.

We present a software suite, named SkyPrism, that implements such 2(3)D modelling for the MAGIC telescope data, including the required computation of a background model and instrument response functions (IRFs). The IRFs are based on Monte-Carlo (MC) simulation and generated during runtime according to the observations; no radial detector symmetry is assumed. We also present the results of validation tests assuring the accuracy of the calculations performed by the package tools. In the summary we discuss how the routines of this package can further be transferred to the analysis of CTA data.
  

\section{The MAGIC SkyPrism package}

The MAGIC SkyPrism package consists of a set of C++/ROOT- and Python-based tools that build on the standard MAGIC data analysis package MARS~\citep{zanin2013mars}. It focuses on the analysis of the entire telescope FoV simultaneously and is designed to work with the 2D images of the sky, possibly binned in energy.

For consistency, all SkyPrism tools (apart from fitting routines) share a single configuration file containing information of the energy binning, data cuts, and the size and binning of the sky image that will be used for the sky, background, and exposure maps. The programs generating the IRFs from the data files, for compatibility reasons, are coded in C++/ROOT, while the routines performing the signal extraction via a likelihood fit are written in Python. We describe the implementation of each of these tools in details below.

%
\subsection{Sky map}

Energy-binned raw sky images, also called ON maps, are 2D histograms in celestial coordinates of all events passing selection criteria. The likelihood analysis routines of SkyPrism will finally fit the model to these ON maps. The events are read from ROOT based event lists \citep{antcheva_root_2009} that are intermediate products of the MAGIC standard data reduction chain of MARS, subjected to selection criteria. They typically consist of the so-called Hadronness cut as well as a lower limit in terms of image size (recorded light yield) in each telescope. The Hadronness parameter of an event is the result of the application of $\gamma$/hadron separation random forests \citep{Magic_RF, breiman2001random}. The same selection criteria as for the ON map are also applied for the events used to generate the instrument responses and background map. 
  
%
\subsection{Pointing history sampling of events for computing the IRFs}
\label{sect:PointHistEvtSamp}

The development of an EAS of a given primary particle type and energy, leaving statistical fluctuations aside, depends primarily on the atmospheric depth in the shower direction and the orientation of the shower axis relative to the Earth's magnetic field. Since for IACTs the atmosphere is an integral part of the detector, the performance of the telescope depends on its pointing direction in the horizontal coordinate system (Azimuth/Zenith). To compute accurate background maps and IRFs, one needs to consider the pointing directions during the observation time. The telescope pointings are weighted with the time spent in the corresponding direction and binned into an Az/Zd histogram, the so-called pointing history, with the Zenith distance (Zd) binned in terms of $\cos$(Zd).\\
To estimate the IRFs of the MAGIC telescopes (also needed in the aperture photometry approach), the MAGIC collaboration generates MC simulated event sets, in which the simulated events are distributed across the entire sky. Out of this event population, the events located around the centre of a filled pointing history bin are sampled to generate the IRFs. The size of the acceptance box around the pointing bin centres can be chosen by the user and by default is the same as the bin size of the pointing history. The method for the background map generation deviates as it requires sampling from measured events, as explained in section \ref{sect:bkg_method}.

%
\subsection{Background map}
\label{sect:bkg_method}

\begin{figure}
	\includegraphics[width=\columnwidth]{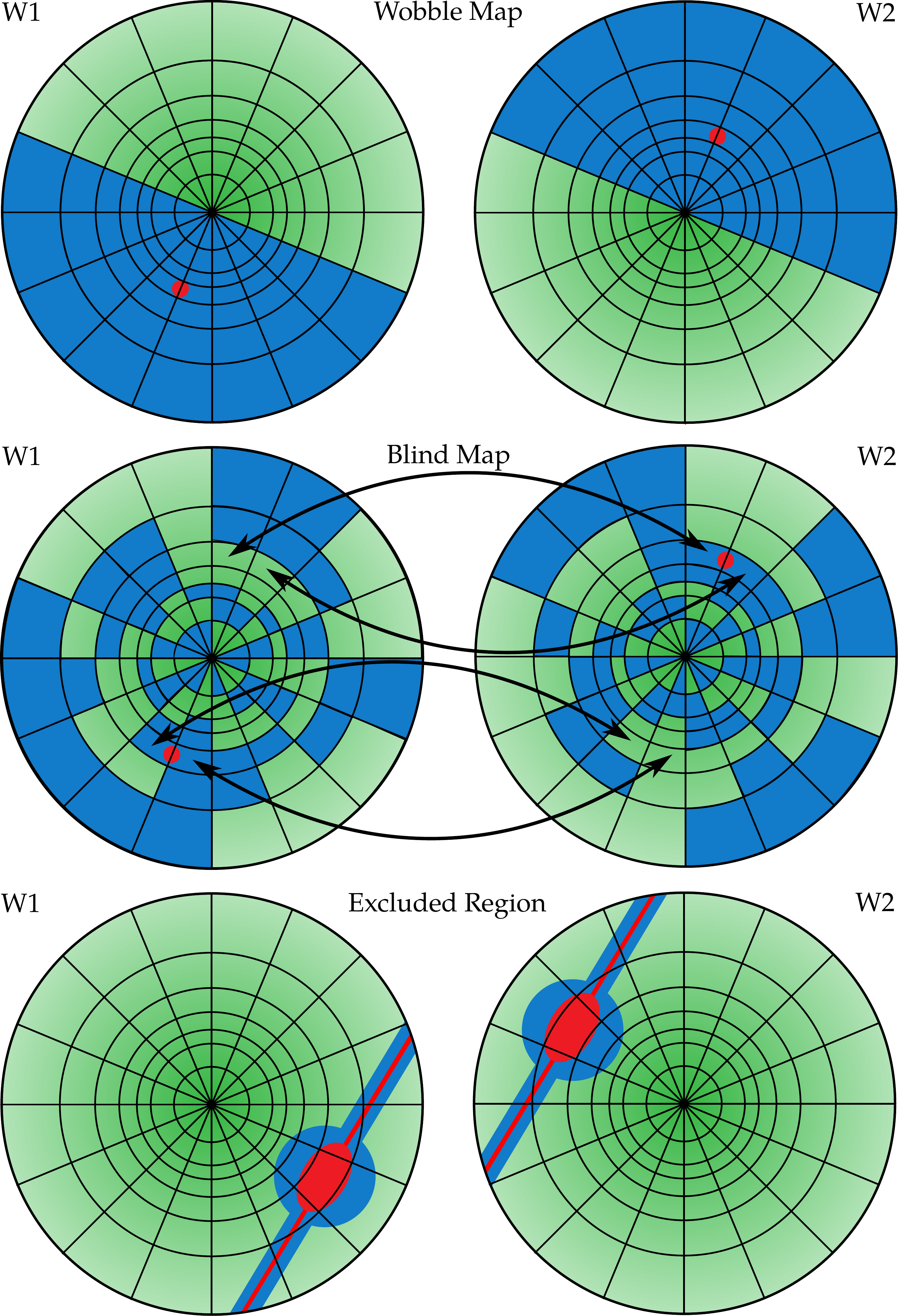}\\
	\caption{Illustration of the different methods for the construction of a background camera exposure model from Wobble observations (here one Wobble pair). The source position and extension is shown as red point, ellipse or stripe. The blue shading marks bins excluded from the background map reconstruction.}
	\label{fig:BkgMethods}
\end{figure}

For weak or extended sources, data collected with IACTs is strongly dominated by an irreducible background~\citep{maier_cosmic_2007, sobczynska09, sobczynska15, sitarek18} that cannot be separated as mentioned in section \ref{sect:introduction}. Therefore, the distribution of background events in the observed sky-patch must be known with high accuracy to keep systematic uncertainties low. Such background counts are caused by the fraction of hadronic CRs surviving the event selection cuts and electron induced showers, which like $\gamma$-ray showers are purely electromagnetic. 

Their distribution in the observed sky region can be expected to resemble the camera response to an isotropic flux of $\gamma$-ray events. For long exposures, however, the background skymap deviates from the simulated $\gamma$-ray exposure. This may result from discrepancies between data and simulations such as dead pixels, stars in the FoV, imperfect flat-fielding of the camera, or a class of hadronic CRs surviving the cuts, but still having an acceptance distribution different to the one of $\gamma$-rays. Hence, one needs to compute the background from measurements.

This background measurement must be preformed under the same observational conditions and follow the same pointing history as the ON data. Such so-called ON-OFF observations are an established mode of observation for IACTs, but are only rarely used due to the additional time needed for the OFF observations and lost for the ON source observations. 

For sources that are point-like or have limited extension ($<\,1/2\,$FoV) the established mode of observation is the so-called wobble mode \citep{1994Fomin_wobble}, where the telescope points in an alternating way to coordinates at a fixed offset around the object. This way ON and OFF measurements can be performed simultaneously since the background can be obtained from another part of the camera of similar acceptance as the source position; the alternation cancels out possible inhomogeneities in the camera halves.

We have implemented three different methods to reconstruct a background exposure model for the whole camera from wobble mode observations. Two of these methods are re-implementations of established techniques \citep[called Wobble and Blind Map;][]{moralejo_mars_2009} used in MARS to generate sky maps. The third method (Exclusion Map) is an advancement of the concept regarding generalization and flexibility. All three methods follow the same basic scheme:
\begin{itemize}
	\item[1.]{All events passing the selection criteria along the pointing track of the telescopes are binned in horizontal coordinates and grouped by Wobble pointings.}
	\item[2.]{A 2D histogram binned in camera coordinates is filled for each Wobble and pointing history bin.}
	\item[3.]{For each pointing history bin the program constructs one camera exposure model out of the different wobble pointings.}
	\item[4.]{The background model in celestial coordinates is populated by random sampling from the corresponding camera exposure models, and applying the correct coordinate rotations and observation time weights along the pointing track.}
\end{itemize}
The methods differ in the way the camera exposures for each wobble pointing are combined into a single assumed source-free background camera exposure model (step 3). The available options illustrated in Fig. \ref{fig:BkgMethods} are:
\begin{itemize}
	\item{{\bf Wobble Map:} the single camera exposures are divided into halves, such that the nominal source position (centre of the wobble setup) is contained in one half, and the normal vector on the separating line is pointing away from this coordinate. The combined background camera exposure model is obtained by normalizing and summing up the source-free halves. This method can only be applied if the $\gamma$-ray emission is confined inside a circle around the nominal source position with a radius corresponding to the Wobble offset.}
	\item{{\bf Blind Map:} the single camera exposures are normalised by the exposure times. From those the combined background camera exposure is obtained by using the median values in each pixel. The median automatically suppresses too large counts from possible sources. This method has the advantage that no prior knowledge about the distribution of $\gamma$-ray sources inside the FoV is needed. However, if only two wobbles are used, the Blind Map by construction underestimates the background, though MARS routines contain a bias correction to compensate for this effect. Also the presence of strong sources will inevitably lead to an upward-bias of the background model in the corresponding regions, causing a systematic flux underestimation for those objects.}
	\item{{\bf Exclusion Map:} here the analyser must specify regions containing known or expected sources, which can have circular or line shape. Those regions are then excluded from the computation of the median as described in the previous method. Special care is required in order not to exclude too large regions, as this might leave no remaining options for the background extraction for certain camera bins.}
\end{itemize}  

%
\subsection{Point spread function}
\label{sect::PSF_method}

IACTs, like all imaging systems, have a finite angular resolution described by the point spread function (PSF). It is defined by the extension of an idealised point-like object, placed at infinity, in the final reconstructed image. Accordingly, the PSF can be seen as the probability function $\mathcal{P}_{\alpha\beta}^{ij}$ for reconstructing a detected event in the bin $\alpha\beta$ instead of $ij$ corresponding to its original arrival direction. 

For MC simulated events the reconstructed and original arrival direction (as reconstructed by a perfect instrument) is known, so one can estimate the angular response. MC events are sampled around the track of the source in the horizontal coordinate system as explained in section \ref{sect:PointHistEvtSamp}. For each event the vector between the reconstructed and original arrival direction in camera coordinates is computed. This vector is finally rotated into the equatorial coordinate system according to section \ref{sect:CoordSysTransformation}. The sum of all events will describe the shape of a point-source observed along the track of the target.

Since the number of MC events is limited, the simulated shape of the PSF is noisy and the PSF is better modelled by a smooth, analytical function. For MAGIC, a 2D double Gaussian~\citep{aleksic_major_2016} or a 2D King~\citep{DAVELA20181} function were found to describe the PSF well. Because of the lower number of free parameters this analysis used the King function:   
\begin{equation}
 K\left(r,\sigma,\gamma\right)=\frac{1}{2\pi\sigma^{2}}\left(1-\frac{1}{\gamma}\right)\left(1+\frac{r^{2}}{2\sigma^{2}\gamma}\right)^{-\gamma}\,,
\end{equation}
where $\sigma$ sets the angular scale of the resulting profile and $\gamma$ determines the weight of the tails. To allow for an asymmetry of the MAGIC PSF, the distance variable $r$ is defined as follows:
\begin{equation}
  \begin{array}{l}
    x_r = ( x_{\alpha\beta} \cos(\phi) - y_{\alpha\beta} \sin(\phi) ) \\
    y_r = ( x_{\alpha\beta} \sin(\phi) + y_{\alpha\beta} \cos(\phi) ) \\
    r = \sqrt{x_r^2+(\epsilon y_r)^2} \,,\\
  \end{array}
\end{equation}
where $\phi$ is a positional angle and $\epsilon$ is the asymmetry parameter, which accounts for the non-symmetry of the MAGIC system~\citep[which has only two telescopes;][]{aleksic_major_2016, DAVELA20181} as well as the influence of the Earth's magnetic field. The King function is fitted to the simulated point-source shape and finally evaluated on a sufficiently large grid around the camera centre.

%
\subsection{Exposure map}
\label{sect:Exposure_method}

The effective area of a telescope corresponds to the size of an ideal detector (with a 100\% detection efficiency) recording the same number of events as the real instrument. It can be defined as the product of the physical detector size and the detector efficiency $\varepsilon_{\text{Det}}$. For IACTs, it is usually estimated with Monte Carlo simulations, where the detection efficiency is a ratio of the detected events to the simulated ones:
\begin{equation}
 A_{\text{eff}} = \frac{N_{\text{det}}}{N_{\text{sim}}} \times \pi \, r_{\text{sim}}^{2}=\varepsilon_{\text{Det}}\times\pi \, r_{\text{sim}}^{2}\,,
 \label{eq:collArea}
\end{equation}
where $N_{\text{det}}$ is the number of detected MC events, $N_{\text{sim}}$ the number of originally simulated events, and $r_{\text{sim}}$ is the maximal simulated distance between the impact point of the shower axis and the telescope (impact parameter), playing the role of an assumed physical detector size.

The Cherenkov light density in the EAS depends on the energy of the primary particle resulting in an energy dependence of the detection efficiency -- and thus the collection area. Additionally, the limited extension of the EAS and the size of MAGIC trigger area~\citep[smaller than the camera size -- has a radius of 1.17$^{\circ}$ vs. 1.75$^{\circ}$;][]{aleksic_major_2016-1}) lead to the drop of the camera acceptance towards its outer regions, resulting in an non-uniform camera response to $\gamma$-ray sources at different distances from the centre of the FoV. All these effects are reproduced in the MAGIC MC simulations.

To correctly estimate the MAGIC detection efficiency from those MC for a specified observation, we follow the procedure outlined in section \ref{sect:PointHistEvtSamp}, but instead of filling the pointing history just with time weights, the effective observation time $T_{\text{eff}}$ for each pointing bin is estimated. It is the sum of the durations of all observational runs in that pointing bin excluding possible gaps of more than 2~seconds and correcting for the instrument's dead time. 

For each pointing bin the program samples events from the MC set and transforms them into the skymap: the program computes the difference between the simulated pointing direction for the telescope and the arrival direction of the event. This vector is added to the equatorial coordinates of the pointing bin (see also appendix~\ref{sect:CoordSysTransformation}). Each event is additionally re-weighted according to the assumed source spectrum with the weights $w(E) = F(E)/M(E)$, where $F(E)$ is the source spectrum and $M(E)$ is the energy spectrum used for the Monte Carlo generation. Due to the limited size of the MC sample, the statistical noise in the obtained maps can exceed the tolerance level of $\sim5\,\%$ required for the accuracy of subsequent source analysis. To overcome this noise, we fit the efficiency model with a modified Gaussian function:
\begin{equation}
   \begin{array}{l}
    x_r = ( x_{ij} \cos(\phi) - y_{ij} \sin(\phi) ) \\
    y_r = ( x_{ij} \sin(\phi) + y_{ij} \cos(\phi) ) \\
    \overrightarrow{r} = \left( s_1 x_r, s_2 y_r \right) \\
    l^2_x = [r-\arctan(r)]_x^2 / (2 \sigma_x^2) \\
    l^2_y = [r-\arctan(r)]_y^2 / (2 \sigma_y^2) \\
    \varepsilon^{\text{Det}}_{ij} = A \times \exp{ (-( l^2_x + l^2_y )) }\,.
  \end{array}
\end{equation}

This parametrisation correctly reproduces the change of the MAGIC detection efficiency throughout the FoV for a wide range of energies, encompassing those usually selected for data analysis. To perform the fit we assume that the number of simulated $n_{ij}$ and detected $k_{ij}$ events in each camera pixel $(ij)$ follows the binomial distribution
\begin{equation}
  P_{ij}\left(k_{ij}|n_{ij},\varepsilon^{\text{Det}}_{ij}\right)= \binom{n_{ij}}{k_{ij}} \left(\varepsilon^{\text{Det}}_{ij}\right)^{k_{ij}} \left(1-\varepsilon^{\text{Det}}_{ij}\right)^{n_{ij}-k_{ij}}\,.
\end{equation}
The best-fit is obtained by minimising the log-likelihood function
\begin{align}
  \ln(\mathcal{L}) = \sum_{ij}k_{ij} \ln{ \left( \varepsilon^{\text{Det}}_{ij} \right) } + \sum_{ij}\left(n_{ij}-k_{ij}\right)\ln{\left(1-\varepsilon^{\text{Det}}_{ij}\right)}\,,
\end{align}
where we have dropped the term $\sum_{ij}\ln{\binom{n_{ij}}{k_{ij}}}$ as it does not depend on the function of interest $\varepsilon^{\text{Det}}_{ij}$. The minimiser used is Minuit2\footnote{\url{https://root.cern.ch/guides/minuit2-manual}} provided by ROOT, which estimates the parameters of the modified Gaussian model and their uncertainties.

The efficiency models, obtained this way, are converted to the collection area via Eq.~\ref{eq:collArea} and, by multiplication with the estimated effective time $T_{\text{eff}}$, to exposure models. This procedure is repeated for each energy bin, selected for analysis in the SkyPrism settings.

To estimate the uncertainties of the camera efficiency models, resulting from the fit, we simulate a number (usually 100) of representations of the model parameters $(A, \phi, s_1, s_1, \sigma_x, \sigma_y)$. The program constructs the parameter sets by drawing random numbers  from a multivariate Gaussian distribution and combining them with the covariance matrix from the best fits. The efficiency models for each of these simulations are stored and can be supplied to the spectral fitting routines of SkyPrism for propagating the fit uncertainties to the flux and spectral parameters estimation.

%
\subsection{Energy dispersion}
\label{sect:Energy_dispersion_matrix}

As well as the reconstructed arrival direction of an event, the reconstructed energy is affected by the detector response and the reconstruction procedure. An event, whose energy falls into a true energy bin $k$, may be wrongly classified into a reconstructed energy bin $l$. This transfer of events between the energy bins can be described by a 2D matrix $\mathcal{D}_{kl}$, called the energy migration or dispersion matrix:

\begin{equation}
\label{equ:energy_migration}
C_\text{obs}\left(E^{\prime}_{l}\right) = \sum\limits_{k} \mathcal{D}_{kl}\left(E^{\prime}|E\right) C_\text{real}\left(E_{k}\right)\,,
\end{equation}
where $C_\text{obs}(E^{\prime}_{l})$ is the measured number of event counts and $C_\text{real}\left(E_{k}\right)$ the count distribution as emitted by the source. For convenience, the migration matrix is normalised along the reconstructed energy axis $E^{\prime}$
\begin{equation}
\sum\limits_l \mathcal{D}_{kl}\left(E^{\prime}|E\right)=1\,,
\end{equation}
which is equivalent to assuming that an event with true energy $E$ will be with 100\% probability contained in the $E^\prime_l$ set.

The energy matrix for a given observation is generated from MC events selected and weighted in the same way as the PSF construction. For the PSF and Exposure program, the user can switch between generating the IRFs in $E$ or $E^{\prime}$.
By default, SkyPrism uses a wider energy range in true energy than that in reconstructed energy, specified by the user. This accounts for the possible spill-over of the events from energy bins outside the analysed energy range to those used in the analysis. By extending the energy range in $E$ down to $1/3\,E^{\prime}_{\text{min}}$ and up to $3\times{}E^{\prime}_{\text{max}}$, we ensure that the lowest and highest true energy bins contribute no more than 10\% to the flux in the $E^{\prime}$ range. Furthermore, the number of bins in $E$ is increased compared to $E^{\prime}$ to allow for an accurate flux reconstruction when changing the spectral parameters during the maximal likelihood fit and to account for possible sharp spectral features such as cut-offs or bumps. By default the binning in the true energy is chosen to approximately match the energy resolution of the MAGIC telescopes.

%
\subsection{Likelihood fitting of the sky maps}
\label{sec:likelihood_fit}

The extraction of the information of the observed sources, such as the flux and the extension, is performed by a set of Python\footnote{https://www.python.org/} routines, which can be accessed from a user script. This way the fit and analysis procedure can be adapted to the user's needs and easily be extended beyond the foreseen functionality presented here.

The fitting procedure combines background, PSF, energy migration matrix and exposure information obtained in the previous steps to provide the maximum likelihood estimate of the fluxes of the sources specified by the user.

Internally, IRFs (e.g exposure and PSF) are represented as 2D images for each of the energy bins where the analysis takes place; they are prepared by the tools described above and loaded at the moment of fitting. The source images are prepared during runtime based on the information, provided in the ``source model'':
\begin{itemize}
 \item for point sources the corresponding image contains one filled skymap bin at the specified source position;
 \item for extended sources the image is taken either from a predefined FITS\footnote{https://fits.gsfc.nasa.gov/fits\_documentation.html}~\citep{FITS} file or a 2D array filled by the user inside the analysis script.
\end{itemize}
Such source maps are further interpolated to exactly match the pixel grid of the sky, background and exposure maps. In each energy bin, the best-fit estimate of the source fluxes is obtained by maximising the Poissonian likelihood of the observed number of counts given the source model and background:
\begin{align}
    \label{eq:Count_model}
    C^{\text{mod}}_{\alpha\beta} &= \sum_{p} \left[\left(\mathcal{S}^{p}_{ij} \times \mathcal{E}_{ij} \right)\otimes \mathcal{P}^{ij}_{\alpha\beta} \right] + \mathcal{B}_{\alpha\beta} \\
    \label{eq:poisson_likelihood_function}
    \mathcal{L} &= \prod_{\alpha\beta} e^{-C^{\text{mod}}_{\alpha\beta}} \frac{\left. C^{\text{mod}}_{\alpha\beta} \right.^{C^{\text{obs}}_{\alpha\beta}}} {C^{\text{obs}}_{\alpha\beta}!}   
\end{align}
$C^{\text{obs}}_{\alpha\beta}$ are the measured counts in each pixel of the sky map, $\mathcal{S}^{p}_{ij}$ the model source ($p$ denotes the source number in the model), $\mathcal{E}_{ij}$ the exposure maps, and $\mathcal{B}_{\alpha\beta}$ the background maps, while $\mathcal{P}^{ij}_{\alpha\beta}$ is the PSF model, mapping an $(i,j)$ pixel to $(\alpha,\beta)$. Computationally the program minimises the negative log-likelihood $-\ln(\mathcal{L})$ using {\tt iminuit}\footnote{http://iminuit.readthedocs.io/en/latest/}. Following the Wilks theorem~\citep{wilks_large-sample_1938}, the uncertainties on the obtained fluxes are computed as deviations of the log-likelihood from its best value by $\Delta \log{L} = \chi^2_1(\alpha)$ (since there is only one parameter of interest -- flux), where $\alpha$ is a desired confidence quantile. Additionally, we have also implemented a Markov-Chain-Monte-Carlo (MCMC) sampling procedure, based on the {\tt emcee}\footnote{http://dfm.io/emcee/current/}~\citep{emcee} library, which allows a more accurate computation of non-symmetric error bars in case of strong correlations between the fit parameters (e.g. if overlapping extended sources are specified in the model).

In addition to the fit of individual source fluxes, the implemented procedure also allows for the source positions to be fitted. For this the supplied source models $\mathcal{S}^{p}_{ij}$ are shifted with respect to the originally specified position; the amounts of the shifts are optimised with the rest of the parameters during the fit. A more sophisticated source position / extension scans can be easily implemented in the user analysis script -- altering the source model and repeating the fit of the already loaded data.

\subsection{Likelihood fitting considering the energy migration}

In addition to fitting of the individual data bins in reconstructed energy, the SkyPrism Python library can also fit in the true energy space via a forward folding procedure. In this case the assumed source spectrum is integrated in all true energy bins $E_{k}$, multiplied with the exposure, and convolved with the PSF corresponding to them. Through a multiplication with the energy migration matrix, the predicted counts of each model component in the reconstructed energy bin $E^{\prime}_{l}$ are obtained. The sum of all components together with the background image (which is always constructed as a function of the estimated energy $E^{\prime}$, as it comes from data directly) leads to the model counts in $E^{\prime}$:
\begin{align}
  \label{eq:Count_model_Etrue}
  C^{\text{mod}}_{\alpha\beta}(E_{l}^{\prime}) = \sum_{kp}  \mathcal{D}_{kl} \left[ \left(\mathcal{S}^{p}_{ij}(E_{k}) \times \mathcal{E}_{ij}(E_{k})\right)\otimes \mathcal{P}^{ij}_{\alpha\beta}(E_{k}) \right] + \mathcal{B}_{\alpha\beta}(E_{l}^{\prime})\,.
\end{align}
The likelihood function in this case is defined as a product of all $E^{\prime}$ bins in Eq.~\ref{eq:poisson_likelihood_function}, which is then optimised with respect to the model parameters via the same routines as described in section \ref{sec:likelihood_fit}.

As opposed to the fits in terms of the estimated energy (Eq.~\ref{eq:Count_model}), the reconstruction of the separate flux points in the true energy space is not a well-defined procedure. It requires assumptions of the source spectral shape between the points (or, equivalently, inside the energy bins, defined in the true energy space). In SkyPrism such differential flux points are interpreted as the nodes of the source spectral model, which uses a linear interpolation of the inter-node energies in the $\log{E} - \log{dN/dE}$ space. This is equivalent to a broken power law with multiple energy break. If the user chooses such a fit during the run time, SkyPrism will use such spectral models for all the sources in Eq.~\ref{eq:Count_model_Etrue} and perform the fit as usual. In this case the energies of the desired data points (nodes) can be chosen freely.

The output of this fit can be used e.g. to assess the validity of the chosen source spectral model. We, however, would like to underline, that the resulting data points (along with their uncertainties) may be strongly correlated and cannot be used as independent measurements in third-party analyses.


\section{Validation on MAGIC data}
To ensure that the SkyPrism tools and routines work as expected, we have performed a series of tests for every part of the package. For the tests, we employ the standard MAGIC Monte-Carlo simulations, that cover all the low-level data reductions and compare the tools outputs with the (a) real data and (b) results of the standard MAGIC data analysis with MARS.


\subsection{Background map}
\label{sect:validation_background}

As the normalization of the background can be refined during the maximum likelihood fit, the background model, described in details in section~\ref{sect:bkg_method}, only needs to match the real background distribution in shape.
In order to estimate a possible mismatch, we have performed
\begin{itemize}
 \item a comparison of the model with the isotropic background, obtained through a simplified simulation of a typical MAGIC observation;
 \item a comparison of the obtained model with the sky map for an observation of an empty region of sky.
\end{itemize}
The simulation has the advantage of eliminating possible small variations of the real instrument performance, providing the consistency check of the procedure itself, whereas the comparison with real data demonstrates the overall performance with all the imperfections included.

\begin{figure*}
	\includegraphics[width=\linewidth]{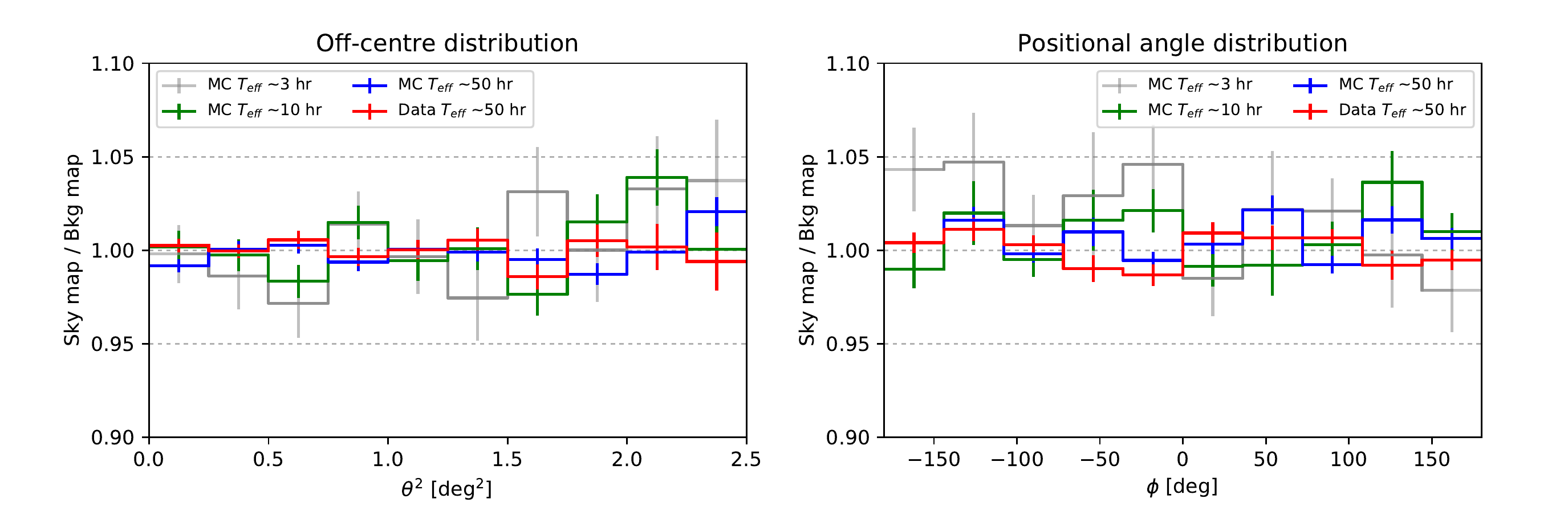}
	\caption{Comparison of a simulated sky map with no source and a background profile, reconstructed from it with SkyPrism. Also shown is a similar profile, obtained from the real empty field data with the BlindMap method of the SkyPrism analysis.
	\textit{Left}: Off-centre distribution.
	\textit{Right}: Positional angle distribution; zero point is arbitrarily set to a horizontal direction.
	}
	\label{fig:Bkg_sim_comparison}
\end{figure*}
For the comparison with the simulated events, we took the MAGIC track on the sky during one of the low zenith angle observations (Zd$\,<35^{\circ}$) and simulated various number of events isotropically distributed over the field of view along the track. The number of simulated events matches to what is usually recorded with the typical analysis criteria in the energy range above 100~GeV with exposure times of 3 to 30~hr. These events were supplied to the Blind map procedure of SkyPrism, with no modifications to the reconstruction settings. 

A comparison of the simulated event distribution with the reconstructed background, obtained in this way, is given in Fig.~\ref{fig:Bkg_sim_comparison}. The left panel of Fig.~\ref{fig:Bkg_sim_comparison} shows the radial distribution of the simulated events with respect to the reconstructed background as function of the offset from the pointing centre. As it can be seen, the performance of the reconstruction method degrades towards the edges of the field of view, where the MAGIC collection area is lower and less events are recorded. Still, spurious background variations, resulting from the SkyPrism methods, do not exceed $\sim 5\%$ even for short exposures ($\sim\,10\,$h). For longer exposures of more than 10 hours, variations are in general less than 2-3\% in the 3.5~deg wide field of view. Similar conclusions can be drawn from the right panel of Fig.~\ref{fig:Bkg_sim_comparison}, depicting the reconstructed background variations as a function of the polar angle with respect to the observation pointing centre.

\begin{figure*}
	\center
	\includegraphics[width=0.8\linewidth]{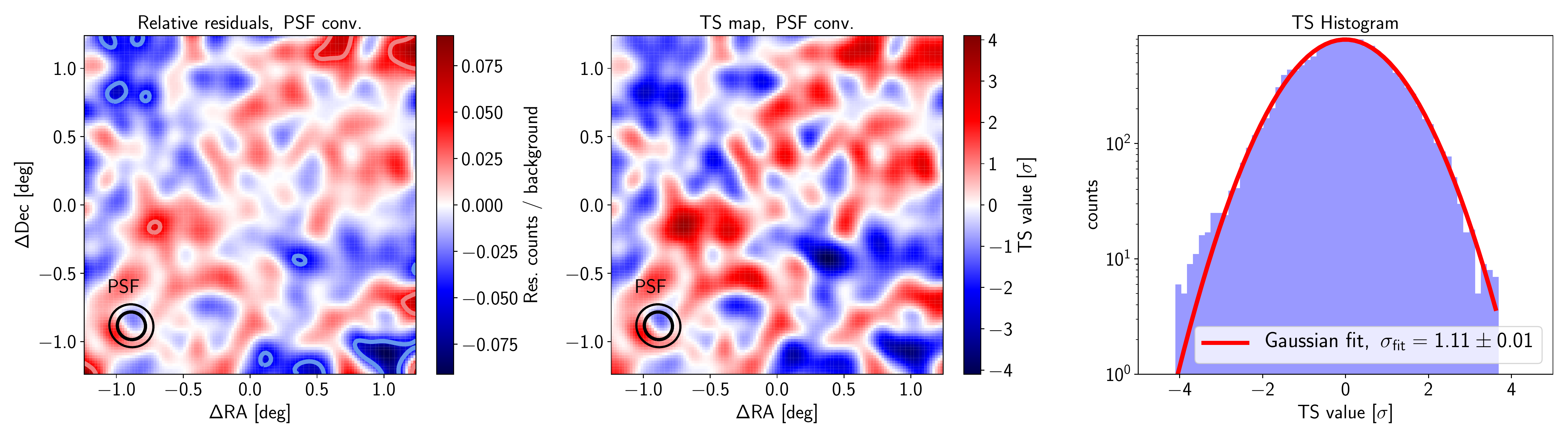}\\
	\includegraphics[width=0.8\linewidth]{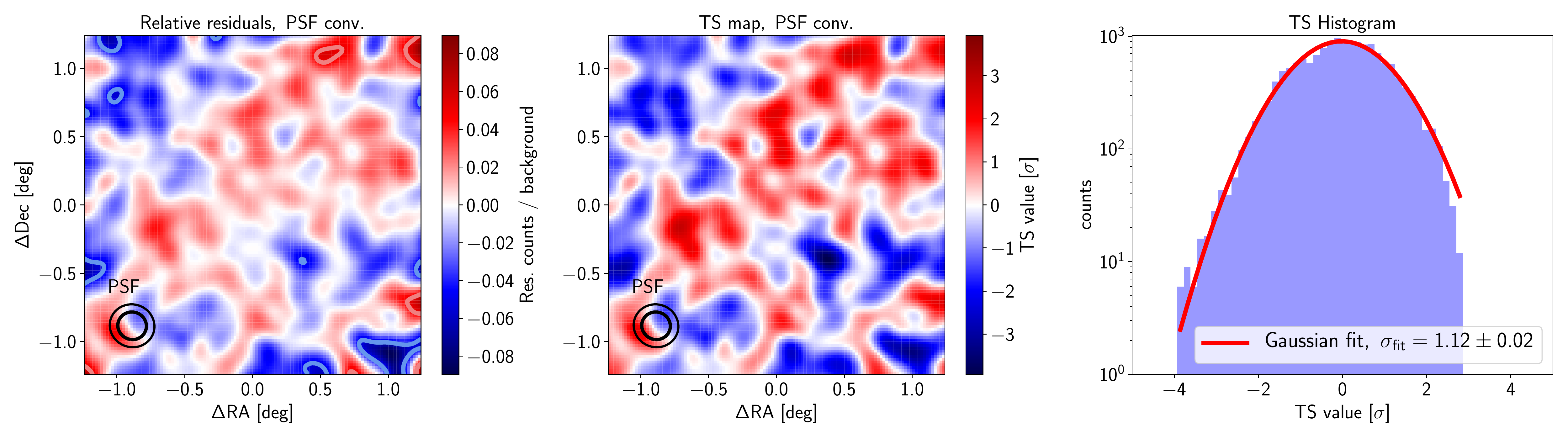}\\
	\includegraphics[width=0.8\linewidth]{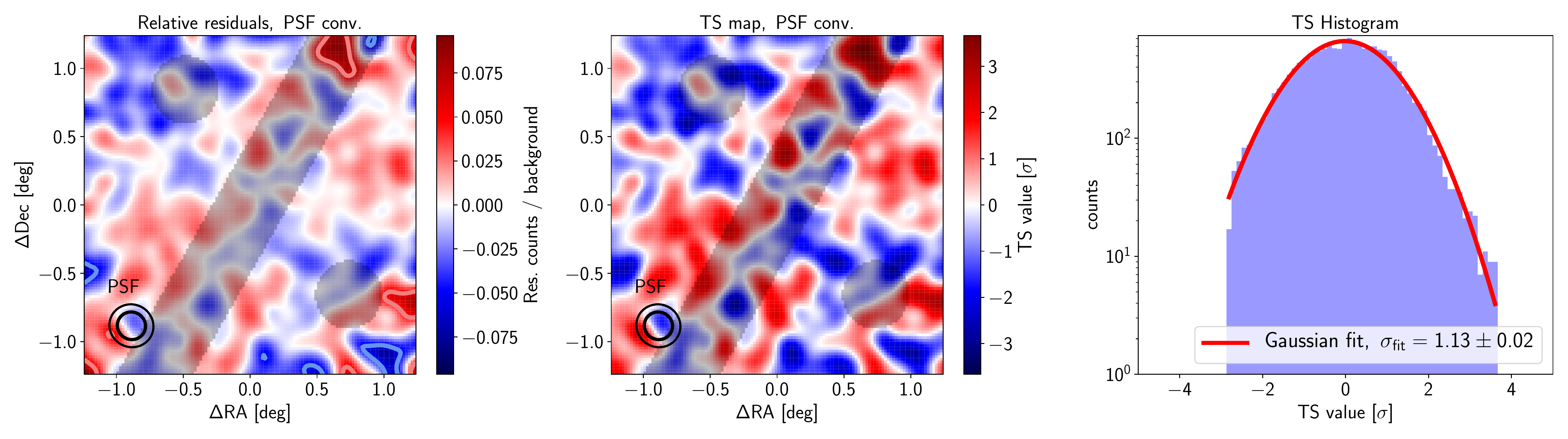}
	\caption{Background methods tested with an empty sky region at Zd$\,<35^{\circ}$. Top to bottom: Wobble map, Blind map, Excluded region map. For the ``Excluded region map'', the ignored regions (a stripe and two circles) are shaded. The PSF after convolution with a smoothing kernel is depicted in terms of 39$\%$ and 68$\%$ containment contours.
	  \textit{Left:} ratio of the residuals (skymap-reconstructed background) to the skymap. The blue and red contours indicate $\pm\,5\%$ relative flux boundary.
	  \textit{Centre:} the test statistic maps computed for the residual counts in the left panels.
	  \textit{Right:} the distribution of the test statistic values (blue histogram) compared to the expected normal distribution (red curve).
	  See Sect.~\ref{sect:bkg_method} and \ref{sect:validation_background} for details.
	}
	\label{fig:Bkg_empty_field_comparison}
\end{figure*}

Additionally, we tested the different background models available within SkyPrism by comparing them to the ON sky map of an empty sky region observed for 50~hr with four wobble pointings. For this purpose, the background model was scaled to match the ON sky map in terms of the median and subtracted from the latter. The residuals were examined in terms of relative flux (residual counts over the background) and test statistic (TS) value. The TS was constructed by sampling 500 times from the background model, using Poissonian random numbers in each pixel. We smoothed the map with a kernel of approximately the size of the PSF to obtain the TS values, roughly equivalent to putative point sources in the field of view. After smoothing, the residuals of the background-subtracted simulations very closely follow a Gaussian distribution around zero in each pixel. We thus define the TS as the local deviation in units of the Gaussian $\sigma$. The results of this test can be seen in Fig.~\ref{fig:Bkg_empty_field_comparison}. As one can see, all three background estimation methods show a similar performance.

The distribution of the TS values matches the normal distribution well enough to not lead to the detections of spurious sources in the image. This can be seen from the the fitted $\sigma_{\text{map}}$ values, given in each TS histogram plot of Fig.~\ref{fig:Bkg_empty_field_comparison}. These values are approximately $1.1 \pm 0.02$, indicating a slight excess on top of the statistical fluctuations alone. This excess is likely due to a small shape discrepancy between the background model and the observed empty field. Assuming that the statistical error $\sigma_{\text{stat}}$ and systematic uncertainty $\sigma_{\text{sys}}$ of the model are both normally distributed, we can quantify the contribution from the systematic component as $\sigma_{\text{sys}}=\sqrt{\sigma_{\text{map}}^{2}-\sigma_{\text{stat}}^{2}}$. This yields $\sigma_{\text{sys}} \approx 0.46$ in units of standard deviation of the simulated background maps. This latter is $\approx 1.8\%$ of the background flux, suggesting that the systematic uncertainty of the SkyPrism background maps is $\sim 1\%$ -- a level comparable to the quoted systematics of the MAGIC telescopes~\citep{aleksic_major_2016}. \newline

It should be noted, that the statistical uncertainty of the constructed background model depends significantly on the event selection criteria. For relaxed cuts and a wide energy range, similar to the ones used here, the relative statistical background uncertainty would scale with the observation time as $\sim 2\%\,\sqrt{50\,\mathrm{h}/t_{\text{obs}}}$.

\subsection{Point spread function}
As mentioned in section \ref{sect::PSF_method}, the MAGIC point spread function in general is not circularly symmetric. This instrumental effect is covered by the MAGIC MCs, but not considered in the standard MAGIC analysis. Instead, the standard MAGIC spectrum extraction tool collects the signal from a circular region around the source, whose extension is usually chosen to contain a significant fraction ($\gtrsim 75\%$) of the signal. This way the positional angle dependence of the PSF has a minor impact on the reconstructed source fluxes in most of the applications of the standard MAGIC analysis chain.

For a 2D image analysis of MAGIC data this effect can not be neglected. For certain observational configurations it would result in noticeable residuals and, consequently, a possible bias in the fit -- especially in crowded regions. To ensure that the effect is properly reconstructed with the SkyPrism tools, we performed a 2D fit of the reconstructed PSF profiles with the King function (see section~\ref{sect::PSF_method} for details), which we compared to the real data.

For this comparison we used a recent MAGIC observation of Mrk~421 at low and medium zenith angles in the energy range from 100~GeV to 10~TeV. Since the PSF calculation procedure, described in section~\ref{sect::PSF_method}, selects only a subsample of available Monte-Carlo events (so that it matches the azimuth and zenith angle distributions of the observations), the number of MC events at the edges of the used energy range is limited. Due to this we dropped the highest (lowest) energy bin for the low (medium) zenith range analysis in order to avoid issues related to the low statistics limit.

\begin{figure*}
	\includegraphics[width=\linewidth]{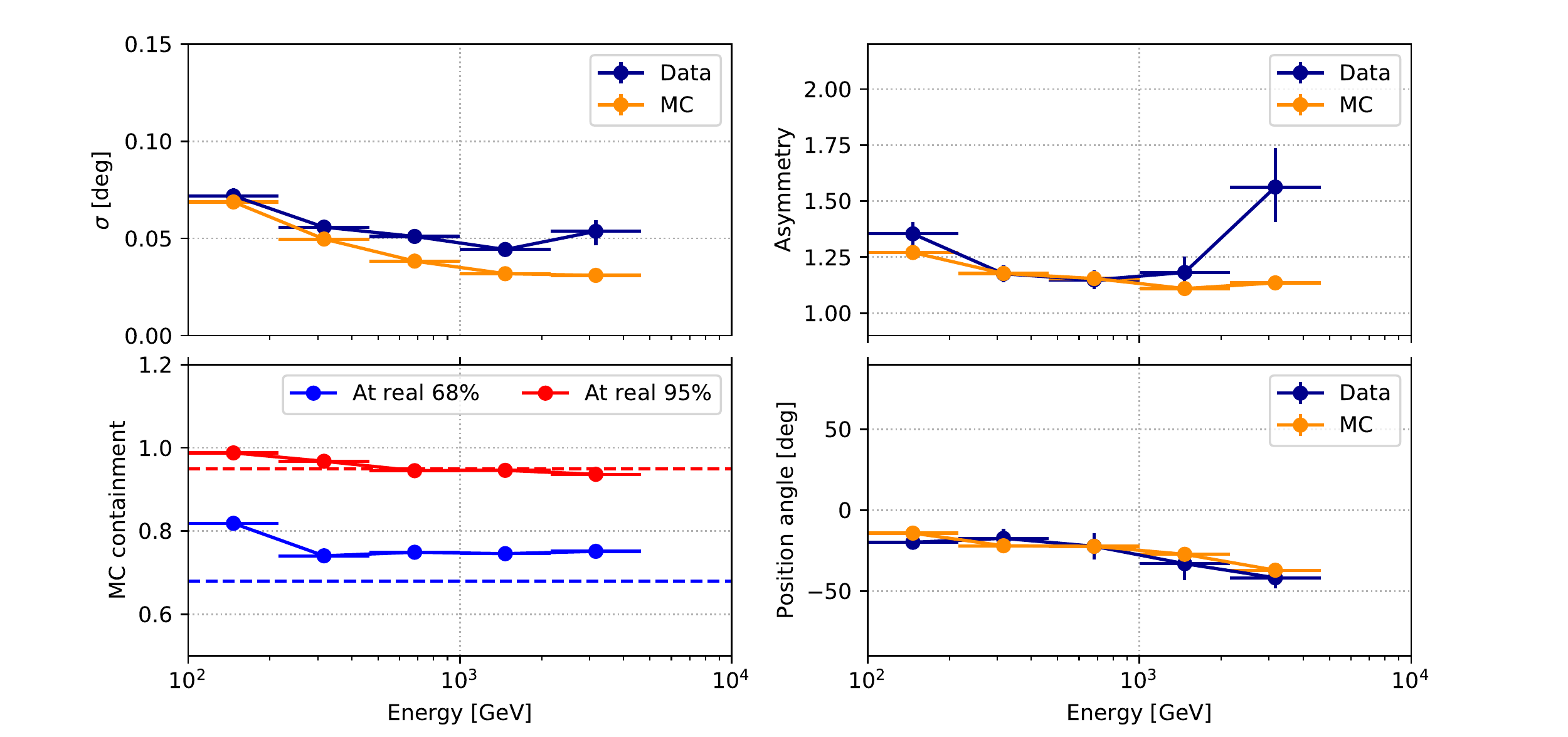}
	\caption{Comparison of the MC PSF to the real data in the $0^\circ-35^\circ$ zenith angle range, performed in terms of the King function fit. 
	\textit{Top left}: $\sigma$ parameter of the King function, which defines the spacial scale.
	\textit{Top right}: asymmetry $\epsilon$ of the fitted profile.
	\textit{Bottom left}: Monte-Carlo events containment, computed at the radial distance corresponding to the 68\% (95\%) data containment radius.
	\textit{Bottom right}: positional angle $\phi$ of the non-symmetric PSF extension.
	}
	\label{fig:PSF_data_comparison_Crab_lowZd}
\end{figure*}

\begin{figure*}
	\includegraphics[width=\linewidth]{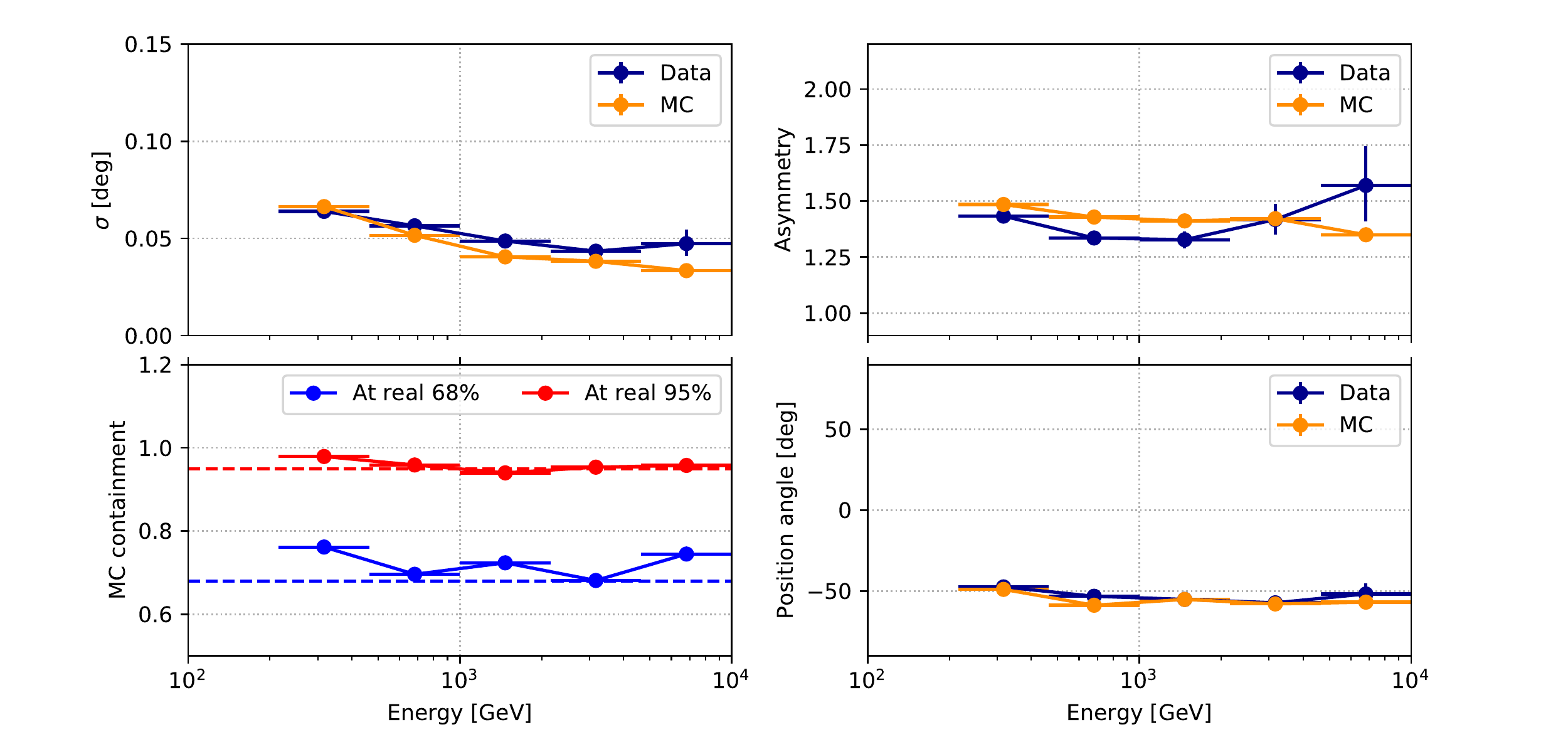}
	\caption{Comparison of the MC PSF to the real data in the $35^\circ-50^\circ$ zenith angle range, performed in terms of the King function fit. Panels are the same as in Fig.~\ref{fig:PSF_data_comparison_Crab_lowZd} 
	}
	\label{fig:PSF_data_comparison_Mrk421_midZd}
\end{figure*}

In Fig.~\ref{fig:PSF_data_comparison_Crab_lowZd} and~\ref{fig:PSF_data_comparison_Mrk421_midZd} we show the comparison of the King function fit results to the low and medium zenith angle data sample as well as to the corresponding MC PSF profile, computed with SkyPrism.
It is evident that the Monte-Carlo estimated PSF is sharper than the real data profile -- the relative event containment difference is~$\sim 10\%$, estimated at the angular distance, where the measured event containment is $68\%$. At larger distances, corresponding to $95\%$ containment of the real data, the Monte-Carlo data mismatch is reduced to~$\lesssim 5\%$. This is comparable to the results obtained in~\citet{aleksic_major_2016}.

\begin{figure*}
	\includegraphics[width=\linewidth]{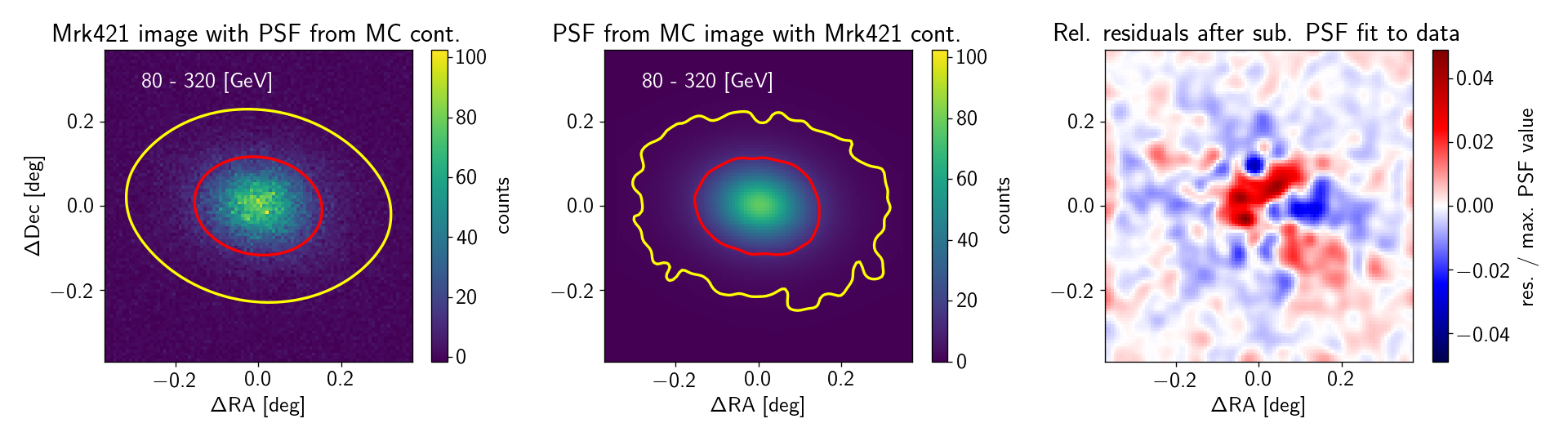}\\
	\includegraphics[width=\linewidth]{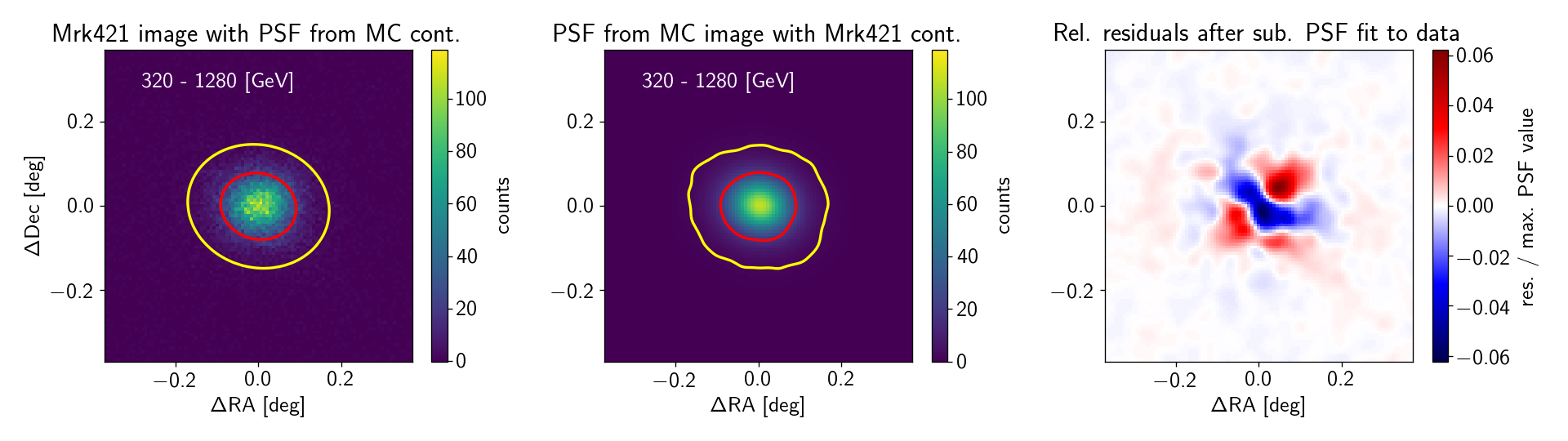}
	\caption{PSF model validation on Mrk421 data taken at zenith angles $<30^\circ$ in two energy ranges: 80~GeV to 320~GeV (top group) and 320~GeV to 1280~GeV) (bottom group). The first column shows the source images with the 68\% and 95\% level contours of the corresponding PSF model; second column gives PSF model with the 68\% and 95\% containment contours of the sky map. The right column shows the residuals after subtracting the normalized PSF model from the sky image.}
	\label{fig:PSF_Check_421}
\end{figure*}

The excessive peakedness of the core of the MAGIC MC PSF was already demonstrated earlier~\citep{aleksic_major_2016}. In practice an additional systematic random component of $\sim 0.02^\circ$ should be added in order to compensate MC to data difference~\citep{aleksic_major_2016}. Indeed, the addition of $0.02^\circ$ smearing to the MC PSF results in $\lesssim 5\%$ residuals in the 2D data-to-model comparison plots, shown in Fig.~\ref{fig:PSF_Check_421}, throughout the entire PSF extension.

Certain small differences between the MC estimated PSF and real data, apparent in the right panels in Figs.~\ref{fig:PSF_data_comparison_Crab_lowZd} and~\ref{fig:PSF_data_comparison_Mrk421_midZd} play a second order role when the MAGIC mispointing is considered. We have checked that the largest deviation in asymmetry for $0^\circ - 35^\circ$ zenith angle (Fig.~\ref{fig:PSF_data_comparison_Crab_lowZd}) range comes from the low number of counts in the source image in the energy bin 2.1\,-\,4.6~TeV. 
At larger zenith angles in Fig.~\ref{fig:PSF_data_comparison_Mrk421_midZd}, the difference in MC and real PSF asymmetries in 215\,-\,464~GeV bin appears significant due to the insufficient number of MC events surviving the selection cuts, which results in a significant underestimation of the corresponding error bars.


Overall, the comparison above demonstrates a reasonable agreement between the Monte-Carlo PSF model, computed with the presented software, and real data.

\subsection{Exposure map}
\label{sect:exposure_validation}

As described in section \ref{sect:Exposure_method}, the exposure and effective area primarily depends on the energy. For a point source our programs should reproduce the same energy dependence of effective area as the standard analysis. As the analysis approach is different, we need to reproduce the aperture photometry approach for estimating the effective area: we cut a circular aperture out of the effective area map and average the effective area inside weighting each bin according to the PSF distribution. This way we obtain the effective area curve for a point source inside a certain area. The size and the data set used are the same as in \citet{aleksic_major_2016}. 

\begin{figure}
	\includegraphics[width=\columnwidth]{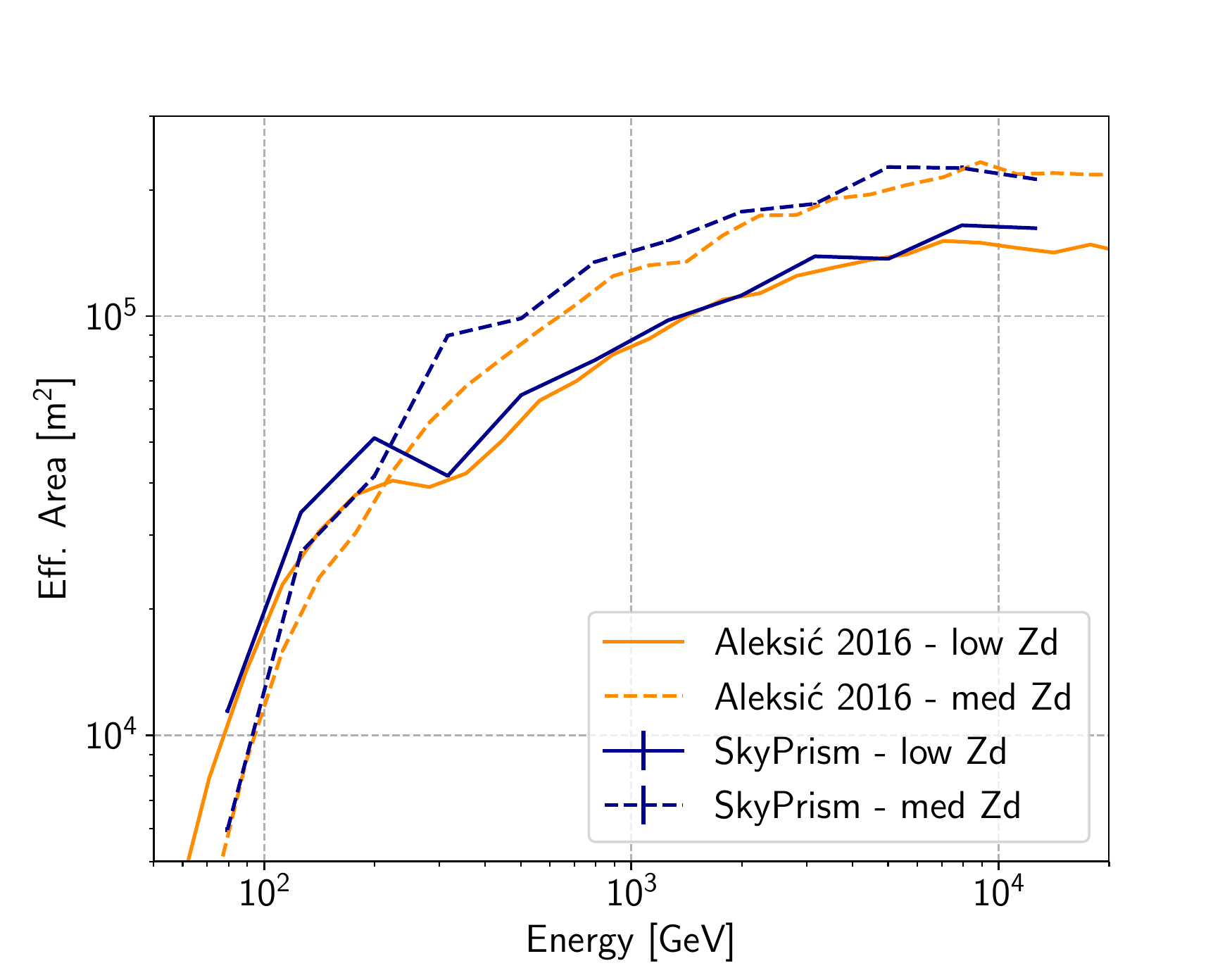}
	\caption{Effective area vs. energy as estimated with the SkyPrism package compared to the standard MAGIC analysis from \citet{aleksic_major_2016}.} 
	\label{fig:Exposure_vs_Energy}
\end{figure}

Figure \ref{fig:Exposure_vs_Energy} shows the comparison of the SkyPrism result with those from \citet{aleksic_major_2016} for two different Zenith distance ranges, low (Zd\,$<\,30^{\circ}$) and medium ($30^{\circ}<$\,Zd\,$<45^{\circ}$). One can see that with higher energy the effective area increases as showers more distant from the telescopes can be recorded. The curves obtained with both methods agree for all energies within 20\%. The discrepancy may well result from the difference in the approaches as they can just be made compatible to a certain extent. Furthermore, the SkyPrism analysis uses Monte-Carlo events distributed over the whole camera, whereas the analysis from \citet{aleksic_major_2016} uses simulation of a point source at an offset of 0.4$^{\circ}$.

As our tools should further compute the effective area correctly across the entire FoV, we validated the estimated off-axis performance. \citet{aleksic_major_2016} measured the rate from the Crab Nebula depending on the offset pointing distance from the source to determine the off-axis acceptance. The measured rate and the effective area off-axis dependence differ only by a single factor, given by the integral spectrum of the source above a considered threshold energy. Since this factor is a constant (for steady sources such as the Crab Nebula), the measured rate and the effective area have the same dependency on the off-axis angle.

\begin{figure}
	\includegraphics[width=\columnwidth]{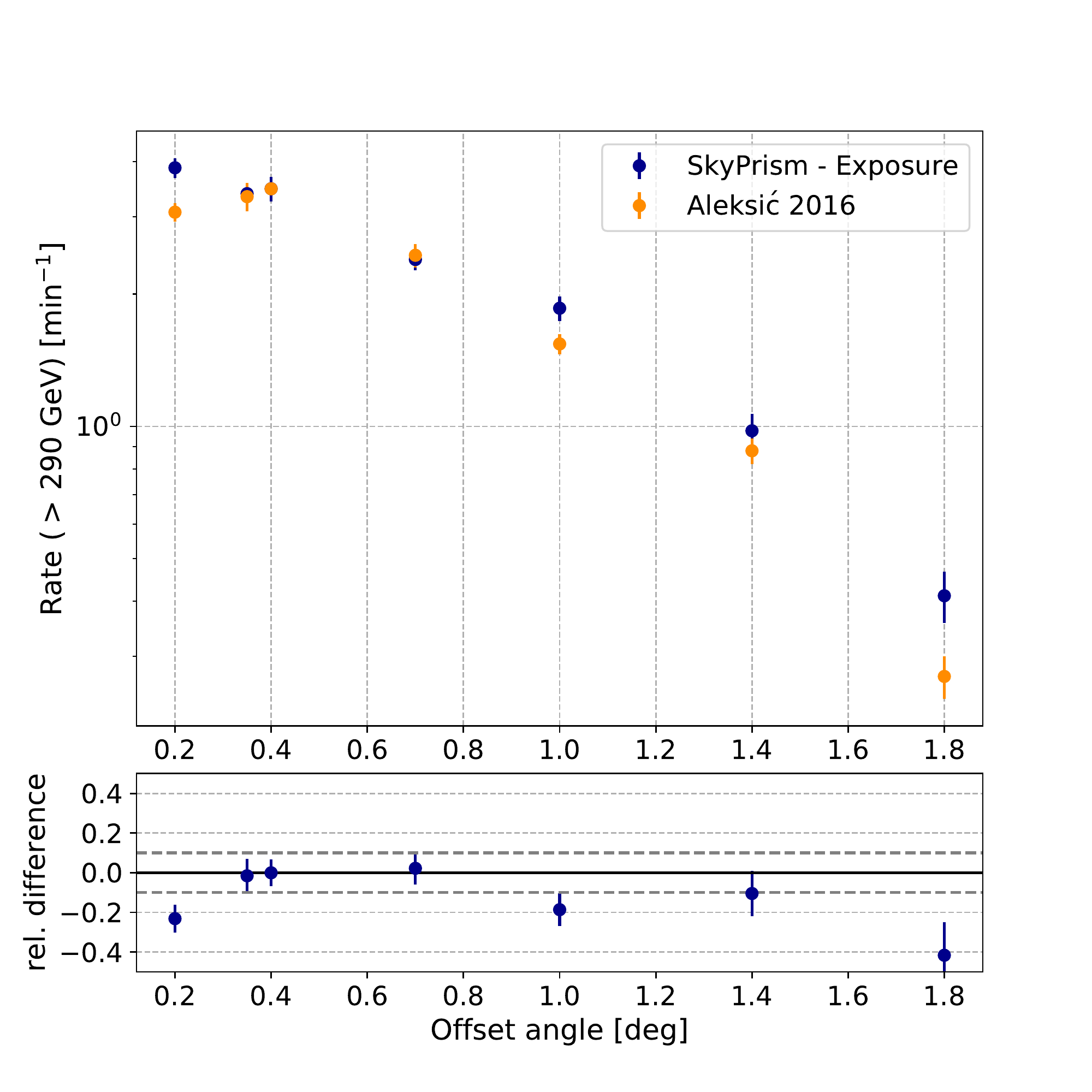}
	\caption{Crab Nebula count rate as a function of the off-axis angle. Orange points correspond to \citet{aleksic_major_2016}, the blue denote the estimates from the SkyPrism exposure model, obtained for the same data and selection cuts. The bottom panel shows the relative difference between both data sets at each offset angle.} 
	\label{fig:Rate_vs_Exposure}
\end{figure}

We computed the effective area for the off-axis data sets of the rate measurement. The effective area was extracted for each off-axis angle separately in the same way as for the energy/effective area comparison. We multiplied the effective area at each offset with the integral spectrum extracted with SkyPrism at 0.4$^{\circ}$ offset to obtain the expected rate. Fig. \ref{fig:Rate_vs_Exposure} shows that the effective area estimates agree with the measurements in nearly all bins within the errorbars, and the relative difference is less than 20\% except for 1.8 deg offset. The MAGIC MC simulation include events up to 1.5 deg offset, hence at 1.8 deg we need to rely on the extrapolation from the fit to the exposure model and thereby the SkyPrism estimation is approximately correct. 

Considering known differences between SkyPrism and the standard MARS analysis, as well as the limited FoV in the MC simulations, the estimations and measurements agree between both approaches. Hence, the SkyPrism package is able to correctly compute the effective area across the MAGIC FoV. 

\subsection{Source flux estimation through the likelihood fit}
\label{sect:fit_validation}

A final test to the overall performance of the SkyPrism package -- and its image fitting routines in particular -- is to reproduce the spectrum of a well-measured source. It is natural to use the Crab Nebula, given that it is among the brightest steady sources in the TeV energy band and is widely used as a calibration object in the IACT community. Here we used the MAGIC Crab Nebula observations at low ($0^\circ - 30^\circ$) zenith angle.

The fit setup for this test includes the PSF, $\gamma$-ray exposure and background model, estimated with SkyPrism, and employs the MAGIC data in the energy range from 60~GeV to 10~TeV. The fitted model includes the point source at the Crab Nebula position and an isotropic background. 

\begin{figure}
	\includegraphics[width=\columnwidth]{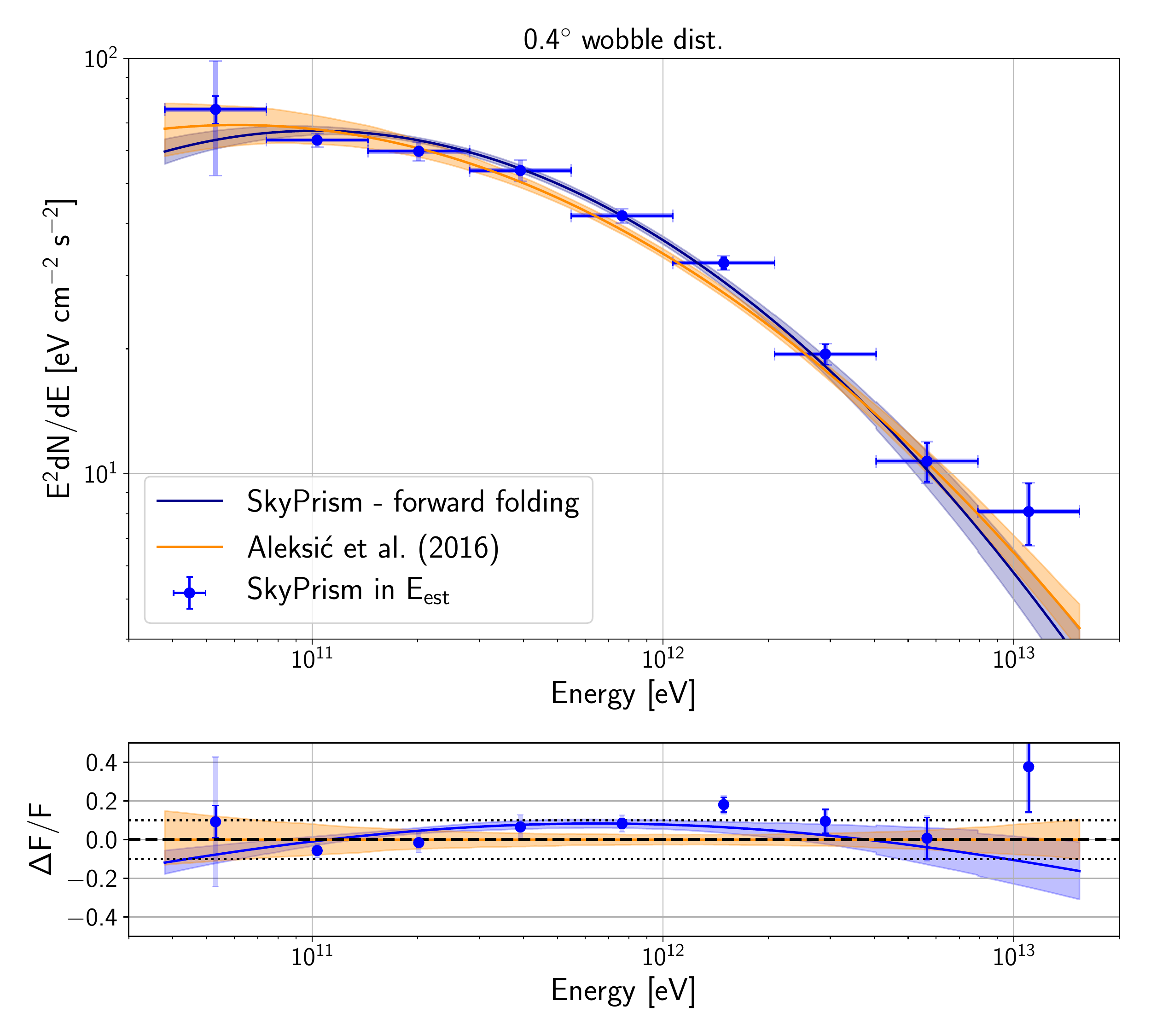}
	\caption{SED of the Crab Nebula obtained by processing the data set used by \citet{aleksic_major_2016} with the full SkyPrism tool chain. The spectral fit is obtained via the forward folding procedure, whereas the data points are the results of the individual fits in each $E_{\text{est}}$ bin. The dark blue error bars of the data points are the statistical error and the light blue ones indicate the uncertainties from the exposure model. The obtained results are also compared to the Crab Nebula SED from the same publication in terms of the relative flux difference ($\Delta{}$F$/$F.)}
	\label{fig:CrabSED_W040}
\end{figure}

The resulting spectrum, presented in Fig.~\ref{fig:CrabSED_W040}, does not show any significant deviations from the reference spectrum, derived by MAGIC~\citep{aleksic_major_2016}. Though the accuracy of the off-axis SkyPrism performance estimate was demonstrated in section~\ref{sect:exposure_validation}, we also performed additional analysis of the larger off-set observations of the Crab Nebula at $0.2^\circ$, $0.7^\circ$, $1.0^\circ$, and $1.4^\circ$ (compared to the standard MAGIC $0.4^\circ$). These tests, presented in Appendix \ref{sec:A}, show the same level of agreement between the standard MAGIC analysis and SkyPrism results, though the amount of data is not sufficient to obtain statistical uncertainties of $\lesssim 10\%$ over the entire energy range.

Overall, this reassures the validity of SkyPrism calculations.

\section{Summary}

We have presented a new IACT data analysis package SkyPrism, optimised for MAGIC data. The package includes several routines that estimate the telescope response (as a function of the field of view and photon energy) from on Monte-Carlo simulations. The estimated response is computed individually for each observation, based on the telescope pointing distribution during the corresponding data taking to ensure it describes the telescope performance accurately.

We have performed extensive tests of the key ingredients of a data analysis: the PSF and $\gamma$-ray exposure, background model and fitting procedure. Altogether we demonstrate an agreement with the standard MAGIC analysis package, MARS~\citep{zanin2013mars}, at the level $\lesssim 10\%$ in terms of the reconstructed source flux. The accuracy of the SkyPrism analysis does not degrade with the off-set from the telescope camera centre, which enables a proper reconstruction of extended sources fluxes over the entire FoV.

With SkyPrism for the first time it is possible to analyse MAGIC data of extended sources of arbitrary morphology. It also can cope with multiple, overlapping point sources in the instrument's field of view, as well as combining several data set covering large sky regions in an convenient way. Though the structure of SkyPrism is different from CTools~\citep{CTools} and Gammapy~\citep{gammapy}, its parts can be further optimised for the data analysis of different IACTs, such as the next-generation CTA.

Within the MAGIC collaboration SkyPrism will be distributed together with the future standard software releases; being open-source, it is also available from the authors upon request.


\begin{appendix}

%
\section{Transformation from Azimuthal to Equatorial Coordinates}
\label{sect:CoordSysTransformation}

While the instrument response depends on the horizontal coordinate system, the emission of astrophysical sources and the background depend on the skyfield, which is preferably defined in a Celestial coordinate system such as the Equatorial coordinate system.


For usual observations the horizontal and equatorial coordinates for the centre of the MAGIC camera are recorded, so the time dependence of the transformation can be factored out. The sampled IRFs can be transformed from horizontal coordinates to equatorial ones via a rotation by an angle $\beta$, which can be computed using the so called navigational triangle: it is formed by the pointing position, the celestial north pole, and the Zenith. As the Azimuth $\gamma$, the latitude of the observatory $\phi$ ($b=90^{\circ}-\phi$), and the Zenith distance of the pointing $a$ are known the rotation angle can be calculated via: 
 
\begin{align}
\label{eq:Rotation}
 c &= \arccos{\left( \cos{(a)}~\cos{(b)} + \sin{(a)}~\cos{(b)}~\cos{(\gamma)}   \right)} \\
 \beta^{\prime} &= \arcsin{\left( \sin{(b)}~\sin{(\gamma)} ~/~ \sin{(c)} \right)} \\
 \beta &= \begin{cases}  \beta^{\prime}  &\mbox{if } \cos{(a)} \cos{(c)} \leq  \cos{(b)} \\
                        \pi - \beta^{\prime} & \mbox{else} \end{cases} 
\end{align}
with c being the distance between the pointing and the celestial north pole.

%

\section{Source spectra estimation at different camera off-set angles}
\label{sec:A}

\begin{figure*}
	\includegraphics[width=\columnwidth]{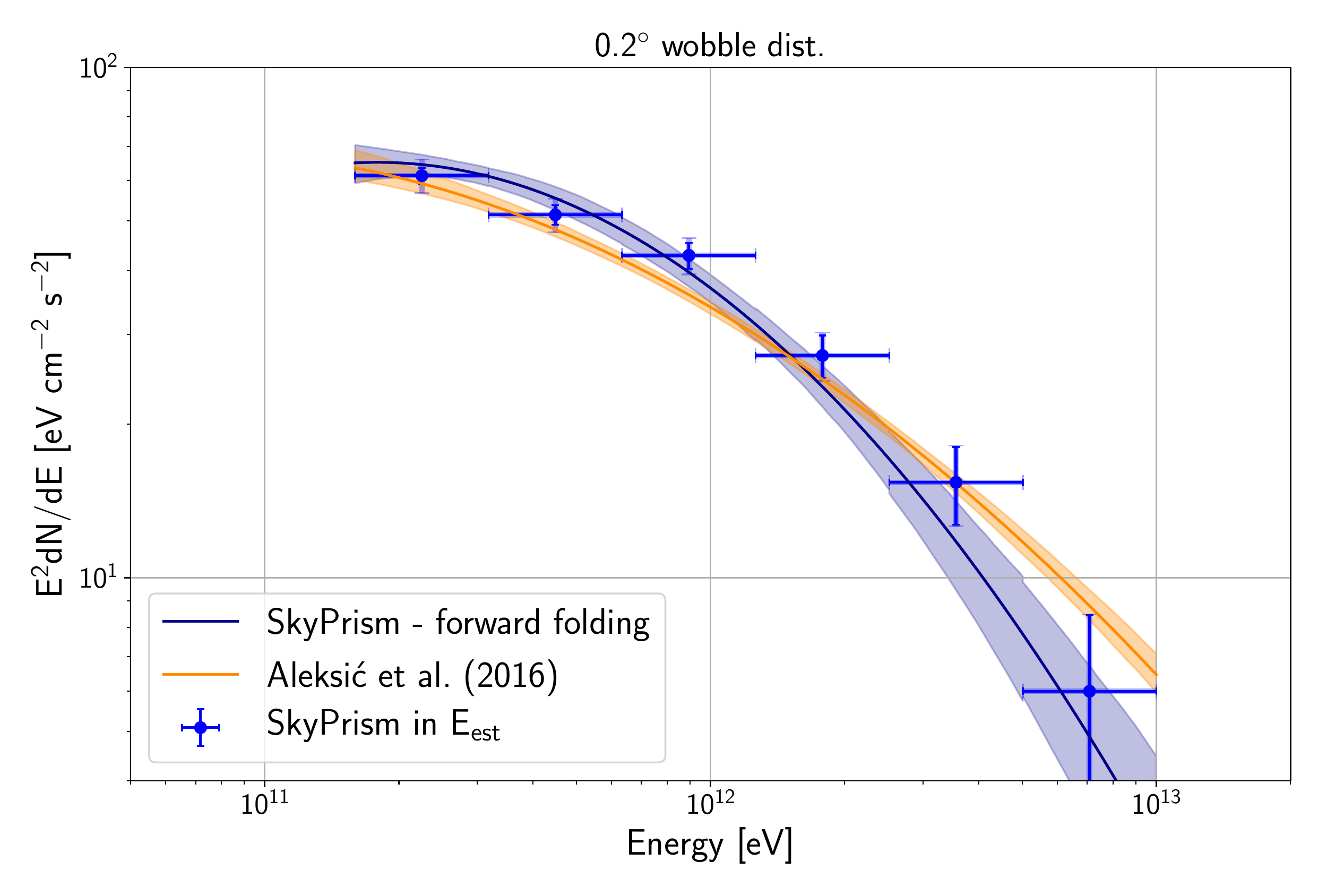}
	\includegraphics[width=\columnwidth]{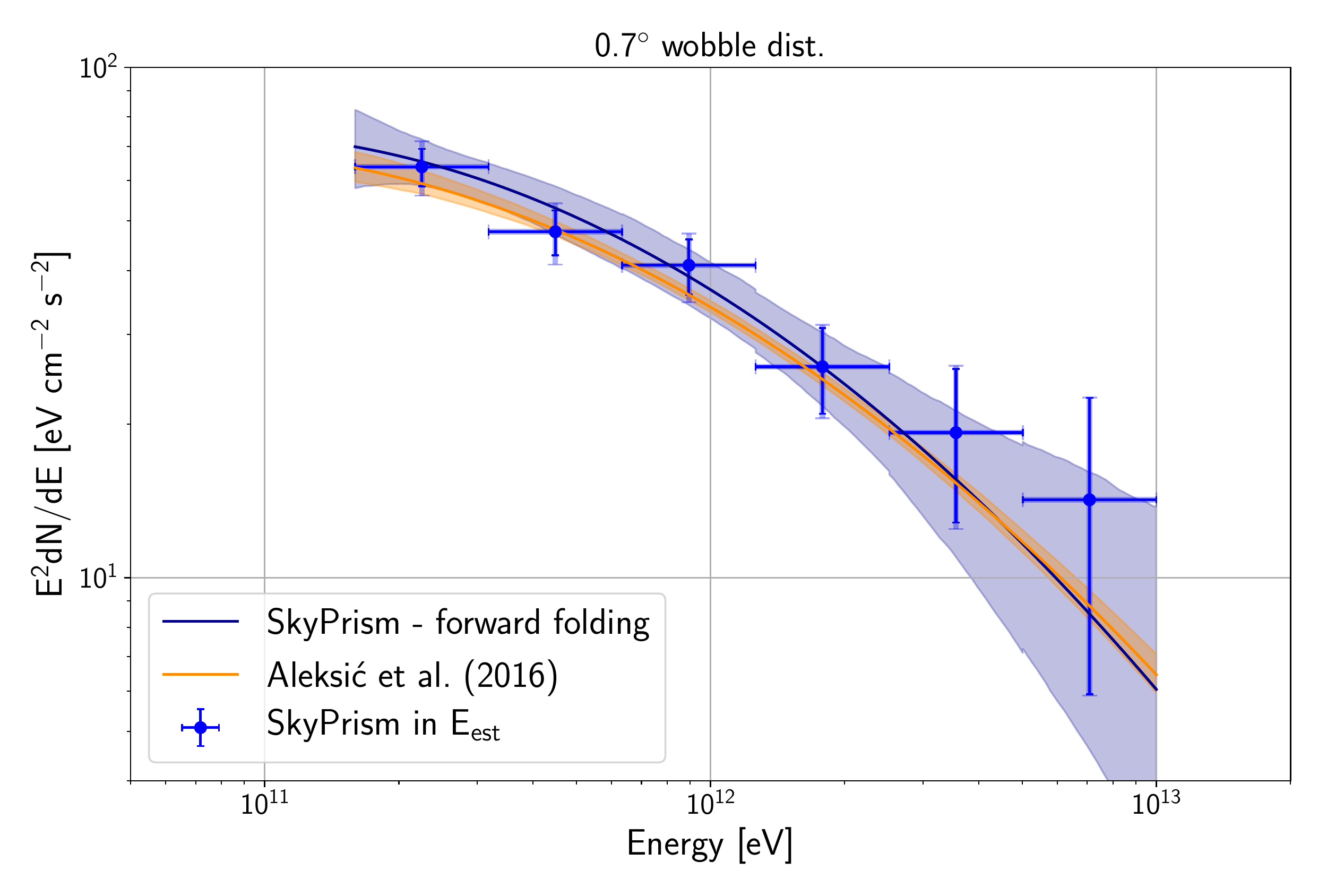}\\
	\includegraphics[width=\columnwidth]{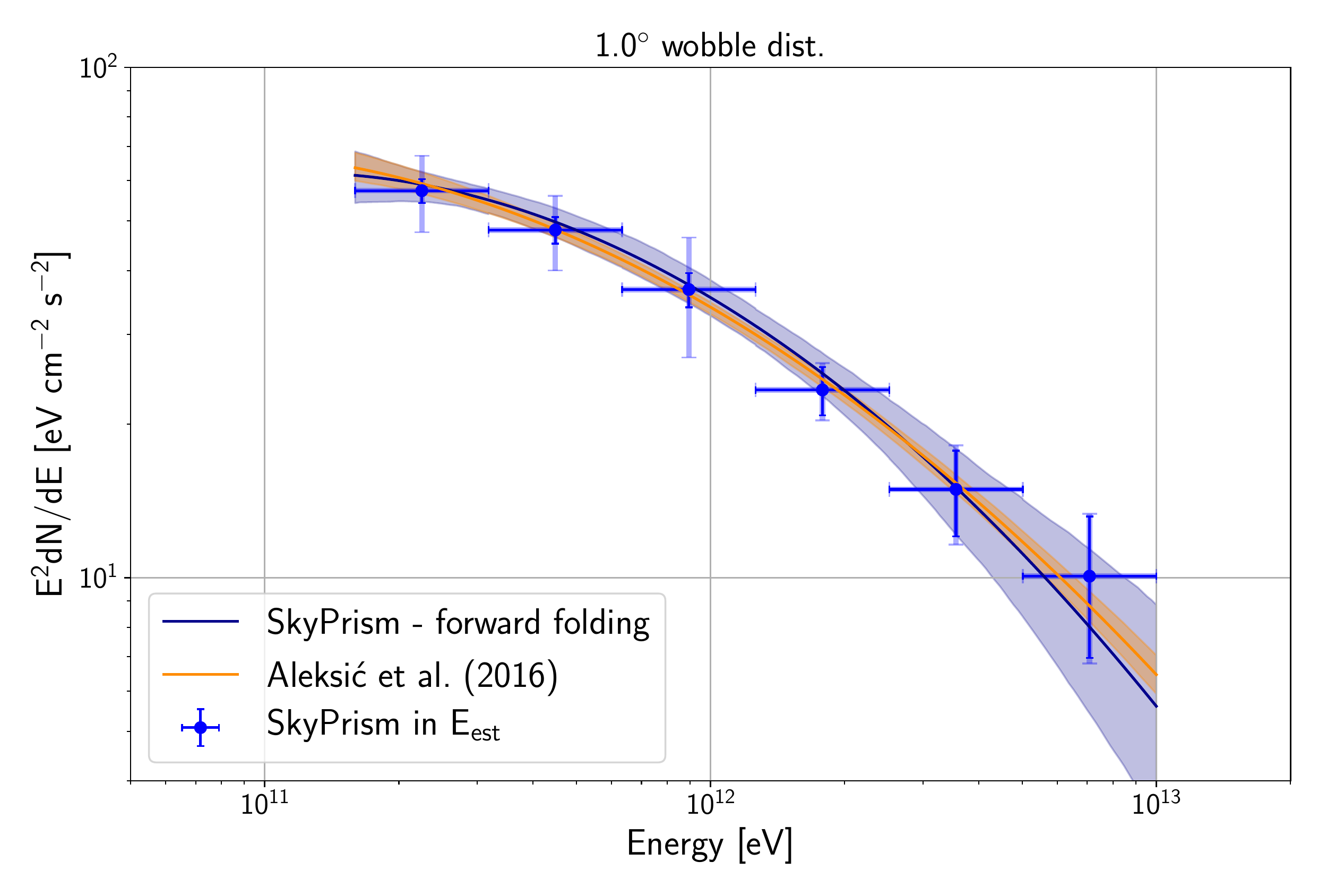}
	\includegraphics[width=\columnwidth]{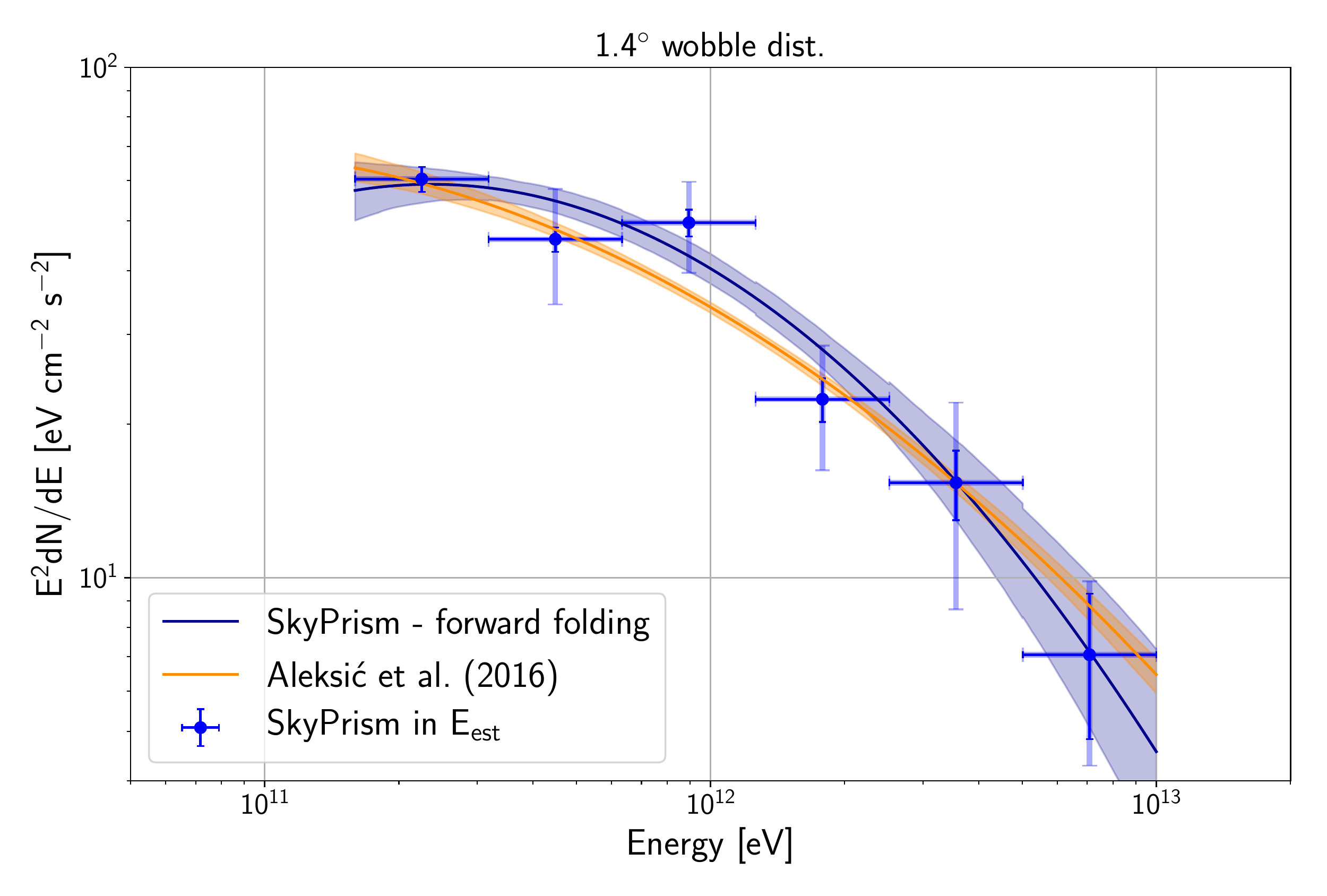}
	\caption{SED of the Crab Nebula obtained by processing data with non-standard wobble offsets that has also been used by \citet{aleksic_major_2016}. The solid error bars indicate statistical errors while the lighter coloured, wider error bars were obtained by also propagating the uncertainties from the constructed exposure model. Due to the short observation times of data sets the uncertainties of the fitted exposure map can be larger than the statistical error of the data points.}
	\label{fig:CrabSED_other_offsets}
\end{figure*}

As a demonstration of the performance of the SkyPrism tools for non-standard source positions with respect to the camera center, here we also show additional SEDs from larger off-set observations of the Crab Nebula, namely at at $0.2^\circ$, $0.7^\circ$, $1.0^\circ$, and $1.4^\circ$ (MAGIC standard off-set is $0.4^\circ$). Similarly to the analysis, presented in Sect.~\ref{sect:fit_validation}, we used the PSF, $\gamma$-ray exposure and background models, constructed with SkyPrism -- for each of the observations individually. The fitted source model again included just the Crab Nebula point source with the addition of the isotropic background. The obtained spectra are shown in Fig.~\ref{fig:CrabSED_other_offsets} and demonstrate a reasonable agreement between SkyPrism and standard MAGIC analysis. The data sets are comparably small, which increases statistical uncertainties.

\end{appendix}


\begin{acknowledgements}
      The authors would like to thank the MAGIC Collaboration for support and providing the data and the MARS software. Particularly, we would like to thank Julian Sitarek for providing us the results of the MAGIC performance paper and valuable discussion throughout the preparation of this work. The work of Ievgen Vovk is supported by the Swiss National Science Foundation grant P2GEP2\_151815.
\end{acknowledgements}



\bibliographystyle{aa}
\bibliography{references}


\end{document}